\documentclass[paper]{JHEP3} 

\usepackage{epsfig}

\newcommand\fverb{\setbox\fverbbox=\hbox\bgroup\verb}
\newcommand\fverbdo{\egroup\medskip\noindent%
			\fbox{\unhbox\fverbbox}\ }
\newcommand\fverbit{\egroup\item[\fbox{\unhbox\fverbbox}]}
\newbox\fverbbox

\newcommand{\mpl}{M_{\rm Pl}}

\newcommand{\del}{\partial}
\newcommand{\phie}{\phih_{\rm{end}}}
\newcommand{\phih}{\phi}
\newcommand{\gh}{g}
\newcommand{\lambdah}{\lambda}
\newcommand{\Mh}{M}

\title{Inflation by non-minimal coupling}

\author{Seong Chan Park and Satoshi Yamaguchi\\
	FPRD, School of Physics and Astronomy, Seoul National University,\\
    Seoul 151-747, Korea\\
	E-mail: \email{spark[at]phya.snu.ac.kr, yamaguch[at]phya.snu.ac.kr}}

\preprint{SNUTP 07-016}	

\abstract{Inflationary scenarios based on simple non-minimal coupling $a \phi^2 R$ and its generalizations are studied.
Generalizing the form of non-minimal coupling to $ K(\phi) R$ with an arbitrary
function $K(\phi)$, we show that the flat potential still is obtainable when $V(\phi)/K^2(\phi)$
is asymptotically constant. Very interestingly, if the ratio of the dimensionless self-coupling constant($\lambda$) of the inflaton field
and the non-minimal coupling constant ($a$) is small, $\sqrt{\lambda /a^2} \sim 10^{-5}$, the cosmological observables for general monomial cases
($K \sim a \phi^m$, $V \sim \lambda \phi^{2m}$) are in good agreement with recent observational data.}


\begin{document}

\section{Introduction}
An early period of accelerated expansion of the universe, or inflation~\cite{inf},
can solve many cosmological problems such as flatness problem, homogeneity problem
and isotropy problem  and can provide the desired initial conditions for the subsequent
hot big bang cosmology~\cite{books}.
In particle physics models, inflation occurs when one or more scalar fields,
the inflaton fields, dominate the energy density of the universe with their
potential being overwhelming~\cite{Lyth:1998xn}. Under such a condition, dubbed
slow-roll condition, the curvature perturbation is produced which is nearly scale
invariant and is heavily constrained by the measurements of the anisotropies of the
CMB and the observations of the large scale structure~\cite{obs}. The slow-roll
condition says that the inflaton potential should be very flat, i.e. the effective
mass of the inflaton should be very small compared with the inflationary Hubble
parameter. The biggest question is the origin of the inflaton field itself.

Recently Bezrukov and Shaposhnikov (BS) reported an interesting possibility that
the standard model with an additional non-minimal coupling term of the Higgs field and
the Ricci scalar ($\sim a |\phi|^2 R$) can give rise to inflation ~\cite{Bezrukov:2007ep}
without introducing any new scalar particle in the theory \footnote{The models of chaotic
inflation with nonzero $a$ were considered in various different contexts
\cite{Spokoiny:1984bd,Salopek:1988qh,Kaiser:1994vs,Komatsu:1999mt,Futamase:1987ua,Fakir:1990eg,Libanov:1998wg}.}.
The authors showed that the ``physical Higgs potential'' in Einstein frame is indeed nearly flat at the large field value limit
but it is also required from the COBE data $U/\varepsilon=(0.027 \mpl)^4$ that the ratio
between the quartic coupling of the Higgs field ($\lambda$) and the non-minimal coupling constant ($a$)
should be small $\sqrt{\lambda/a^2}\sim 10^{-5}$ \footnote{One should note that when one is trying to identify the inflaton field as the Higgs field the self-coupling $\lambda$ is of the order of unity. In that case the largeness of the non-minimal coupling is required as well.}.
Here we would generalize the case of BS by taking more generic form of the nonminimal coupling and read out the required condition for
the asymptotically flat potential. It is certainly worthwhile to consider the generalization since we could understand the underlying structure of
the theory more closely.

After reviewing the suggestion by BS in the next section, we generalize
the suggestion by considering generic form of the gravity-scalar coupling term in a non-minimal way ($K(\phi)R$) and see the general
condition for getting the flat potential or the slow-roll condition in the Sec.\ref{generalization}. In the Sec.\ref{monomial},
we work out the monomial case with functions ($K\sim a \phi^m$ and  $V\sim \lambda \phi^{2m}$) in detail. Interestingly for any positive integer power ($m$) the scalar spectral index and the tensor-to-scalar perturbation turned out to be in good agreement with the latest cosmological observations
once a combination of dimensionless self coupling constant and the coefficient of the nonminimal coupling is fixed as $\sqrt{\lambda_0/a_0^2}\sim 10^{-5}$ (Details of the parameters are given in Sec.\ref{monomial}). Summary and discussions will be followed.

\section{Review: The Higgs boson as the inflaton}
\label{review}
The starting action functional for the gravity-Higgs scalar system in the Jordan frame is given as follows.
\begin{eqnarray}
S_J = \int d^4 x \sqrt{-g} \left[-\frac{M^2+ a \phi^2}{2}R + \frac{(\partial \phi)^2}{2}-V(\phi)\right]
\end{eqnarray}
where the scalar potential is $V(\phi) = \lambda/4 (\phi^2 -v^2)^2$ and $M\simeq \mpl$ in the parameter regime
$1\ll a \ll (\mpl/\langle\phi\rangle)^2$ assuming the vacuum
expectation value $v$ is in the electroweak scale. One can move to the Einstein frame where the graviton
kinetic term is canonical by the conformal transformation
\begin{eqnarray}
g_{\mu\nu}\rightarrow g^E_{\mu\nu} = e^{2 \omega} g_{\mu\nu},
\end{eqnarray}
where $g^E$ is the metric in the Einstein frame and the conformal factor is defined as $e^{2\omega} = 1+ a \phi^2/\mpl$.
The resultant action in the Einstein frame provides the ``Higgs potential''
\begin{eqnarray}
U(h(\phi)) = e^{-4 \omega} V(\phi)
\label{potential}
\end{eqnarray}
where the canonically normalized scalar field $h$ is determined from the scalar field $\phi$ by a derivative:
\begin{eqnarray}
\frac{d h }{ d\phi} = \sqrt{\frac{e^{2\omega}+ 6 a^2 \phi^2/\mpl^2}{e^{4 \omega}}}.
\end{eqnarray}
Here the coefficient `6' comes from the conformal transformation of the Ricci scalar, $R\rightarrow e^{-2\omega}(R-6 \nabla^2 \omega - 6(\partial \omega)^2)$ \footnote{In $D$ dimensions, $R\rightarrow e^{-2\omega}\left(R- 2(D-1)\nabla^2\omega-(D-2)(D-1)(\partial \omega)^2\right)$.}.
Taking $h \gg \mpl/\sqrt{a}$, the potential in eq.\ref{potential} can be recasted as
\begin{eqnarray}
U(h) &\simeq& \frac{\lambda \mpl^4}{4 a^2}\left(1+ e^{-2 h/(\sqrt{6}\mpl)}\right)^{-2} \nonumber \\
&\simeq&  \frac{\lambda \mpl^4 }{4 a^2}.
\label{potential in Einstein}
\end{eqnarray}
i.e. the potential becomes very flat at high energy regime. Bezrukov and Shaposhnikov found that if the coefficient of
the non-minimal coupling $a$ is chosen {\it properly}, this potential really reproduce the current data from
the cosmological observations. The value is $\sqrt{\lambda/a^2} = 2.1 \times 10^{-5}$. As seen in eq.\ref{potential in Einstein},
the flat potential is not directly related with the actual vacuum expectation value at low energy. Indeed, inflation
takes place at the high energy regime which should be independent of the details of the low energy values.

\section{Generalization: $ a \phi^2 R \rightarrow K(\phi) R$}
\label{generalization}

We generalize the model with non-minimal coupling $K(\phih)$, and the scalar potential $V(\phih)$. The starting 4-dimensional action in Jordan frame is expressed as
\begin{eqnarray}
 S=\int d^4 x \sqrt{-\gh}\left[-\frac{\Mh^2+K(\phih)}{2}{R}+\frac{1}{2}(\del \phih)^2-V(\phih)\right].
\end{eqnarray}
The Einstein metric is
\begin{eqnarray}
 \gh_{\mu\nu}=e^{-2\omega}g^{E}_{\mu\nu},\qquad
 e^{2\omega}:=\frac{\Mh^2+K(\phih)}{\mpl^2}.
\end{eqnarray}
Then the action in the Einstein frame becomes
\begin{eqnarray}
 \int d^4x\sqrt{-g_{E}}\left[
-\frac{\mpl^2}{2}R_{E}
+\frac34\frac{e^{-4\omega}}{\mpl^2}K'(\phih)^2(\del \phih)^2
+\frac12 e^{-2\omega}(\del \phih)^2
-e^{-4\omega}V(\phih)
\right].
\end{eqnarray}
It is convenient to redefine the scalar field and normalize the kinetic term canonically.
\begin{eqnarray}
 \frac{dh}{d\phih}=\sqrt{
\frac{\mpl^2}{\Mh^2+K(\phih)}
+\frac32\frac{\mpl^2}{(\Mh^2+K(\phih))^2}K'(\phih)^2
}.\label{dhdphi}
\end{eqnarray}
Now the scalar potential is written as
\begin{eqnarray}
 U=\frac{\mpl^4}{(\Mh^2+K(\phih))^2}V(\phih). \label{U}
\end{eqnarray}
Here we could read out the general condition for the flat potential
at the large field value:
\begin{eqnarray}
\lim_{\phih\rightarrow \infty}\frac{V}{K^2} = Const >0.
\label{condition}
\end{eqnarray}
since $U \varpropto \frac{V}{K^2}$. The condition $K(\phih) \gg \Mh^2$ for $\phih \gg \Mh$
is required for the potential to be bounded from below and the location of the global minimum
is well localized around the small field value.

Here we would like to add one comment about the condition.
Even though the condition in eq. \ref{condition} actually determines the flatness of the potential at the large field value,
it is not necessarily required in generic inflation models. Depending on the shape of the potential, it might still be
possible to have sufficient time of exponential expansion for some {\it finite} region of field value $\phi$.
The result is applicable for monotonic potentials, for example, monomial potentials which will be considered
below in great detail.

\section{Monomial case: $K\sim \phi^m$}\label{monomial}

Let us take $K(\phih)$ to be a monomial as
\begin{eqnarray}
 K(\phih)=a \phih^m,
\end{eqnarray}
where $a$ is a dimensionful constant.
In order to get the flat potential in large $\phih$ region in Einstein frame, the original scalar potential in Jordan frame should be written as
\begin{eqnarray}
 V=\frac{\lambdah}{2m}\phih^{2m}.
\end{eqnarray}
In this case, $U$ is written as
\begin{eqnarray}
 U=\frac{\mpl^4\lambdah}{2m a^2}\left(1+\frac{\Mh^2}{a}\phih^{-m}\right)^{-2}
 \label{potentialm}
\end{eqnarray}

In large $\phih$ region, the relation \ref{dhdphi} between $\phih$ and $h$ is written as
\begin{itemize}
 \item $m=1$
\begin{eqnarray}
 \frac{dh}{d\phih}\cong\frac{\mpl}{\sqrt{a}}\frac{1}{\sqrt{\phih}},\qquad
 \phih\cong \frac{a}{4\mpl^2}h^2. \label{hphi1}
\end{eqnarray}
 \item $m=2$
\begin{eqnarray}
 \frac{dh}{d\phih}\cong\sqrt{6+1/a}\frac{\mpl}{\phih},\qquad
 \phih\cong \frac{\mpl}{\sqrt{a}}\exp\frac{h}{\sqrt{6+1/a}\mpl}.\label{hphi2}
\end{eqnarray}
 \item $m\ge 3$
\begin{eqnarray}
 \frac{dh}{d\phih}\cong\sqrt{\frac32}\frac{m \mpl}{\phih},\qquad
 \phih\cong \left(\frac{\mpl^2}{a}\right)^{1/m}\exp\sqrt{\frac23}\frac{h}{m \mpl}.\label{hphi3}
\end{eqnarray}
\end{itemize}

The slow roll parameters are defined by using the scalar potential in Einstein frame \ref{U} and the canonically normalized scalar field $h$ as
\begin{eqnarray}
 \varepsilon=\frac{\mpl^2}{2}\left(\frac{\del U/\del h}{U}\right)^2,
 \qquad
 \eta=\mpl^2\frac{\del^2 U/\del h^2}{U}.
\end{eqnarray}

In our model these parameters are calculated in large $\phih$ region, using eqs.
\ref{potentialm}, \ref{hphi1}--\ref{hphi3}, as
\begin{eqnarray}
 \varepsilon=
 \left\{
   \begin{array}{ll}
     \frac{2M}{a}\left(\frac{M}{\phih}\right)^3, & \hbox{$m=1$;} \\
     \frac{4}{3a^2(1+1/(6a))}\left(\frac{M}{\phih}\right)^4, & \hbox{$m=2$;} \\
     \frac{4M^{-2m+4}}{3a^2}\left(\frac{M}{\phih}\right)^{2m}, & \hbox{$m\ge 3$.}
   \end{array}
 \right.
,
 \eta=
 \left\{
   \begin{array}{ll}
     -3\left(\frac{\Mh}{\phih}\right)^2, & \hbox{$m=1$;} \\
     -\frac{4}{3a(1+1/(6a))}\left(\frac{\Mh}{\phih}\right)^2, & \hbox{$m=2$;} \\
     -\frac{4\Mh^{2-m}}{3a}\left(\frac{\Mh}{\phih}\right)^m, & \hbox{$m\ge 3$.}
   \end{array}
 \right.
\label{epsilon-eta}
\end{eqnarray}

The end of inflation is $\varepsilon=1$. The values of $h$ and $\phih$ at this point are denoted by $h_{\rm end}$ and $\phie$ respectively. In the slow roll inflation the number of e-foldings is expressed as
\begin{eqnarray}
 N=\frac{1}{\mpl^2}\int^{h_0}_{h_{\rm{end}}}\frac{U}{\partial U/\partial h}.
\end{eqnarray}
In our model $N$ is calculated as
\begin{eqnarray}
 N=
\left\{
\begin{array}{ll}
 \frac{1}{4M^2}(\phih_0^2-\phie^2), & \hbox{$(m=1)$}\\
 \frac{3}{4} a\left(1+\frac{1}{6a}\right)\frac{1}{M^2}(\phih_0^2-\phie^2), & \hbox{$(m=2)$}\\
 \frac{3}{4} a\frac{1}{\Mh^{2}}(\phih_0^m-\phie^m) , & \hbox{$(m \ge 3)$}
\end{array}
\right.
\end{eqnarray}
In order to get $60$ e-foldings, we should solve $N=60$ and get $\phih_{60}$.
Let us assume $\phih_{60} \gg \phie^2$. Then we obtain the value $\phih_{60}$ as
\begin{eqnarray}
  \phih_{60}=
\left\{
  \begin{array}{ll}
    2\sqrt{N}M, & \hbox{($m=1$)} \\
    \frac{2 \sqrt{N}M}{\sqrt{3a(1+1/(6a))}}, & \hbox{($m=2$)} \\
    \left(\frac{4 N}{3a}M^2\right)^{1/m}, & \hbox{($m\ge 3$).}
  \end{array}
\right.
\label{phi60}
\end{eqnarray}

The spectral index $n_s$ and the tensor-to-scalar ratio $r$ can be calculated as
\begin{eqnarray}
 n_s=1-6\varepsilon+2\eta|_{\phih=\phih_{60}},\qquad
 r=16\varepsilon|_{\phih=\phih_{60}}.
\end{eqnarray}
In our model, these values are expressed (using eq.\ref{epsilon-eta} and eq.\ref{phi60}) as
\begin{eqnarray}
 n_s=
\left\{
\begin{array}{ll}
 1-\frac{3}{2a_0N^{3/2}}-\frac{3}{2N}, & \hbox{$(m=1)$}\\
 1-\frac{9(1+1/(6a_0))}{2N^2}-\frac{2}{N}, & \hbox{$(m=2)$}\\
 1-\frac{9}{2N^2}-\frac{2}{N}, & \hbox{$(m\ge 3)$}
\end{array}\right.
,\qquad
r=
\left\{
\begin{array}{ll}
 \frac{4}{a_0N^{3/2}}, & \hbox{$(m=1)$}\\
 \frac{12(1+1/(6a_0))}{N^2}, & \hbox{$(m=2)$}\\
 \frac{12}{N^2} , & \hbox{$(m\ge 3)$}
\end{array}\right.
\end{eqnarray}
where the dimensionless parameter $a_0$ is defined as
\begin{eqnarray}
 a_0=a M^{m-2}.
\end{eqnarray}

\begin{figure}\begin{center}
\includegraphics[width=0.65\textwidth]{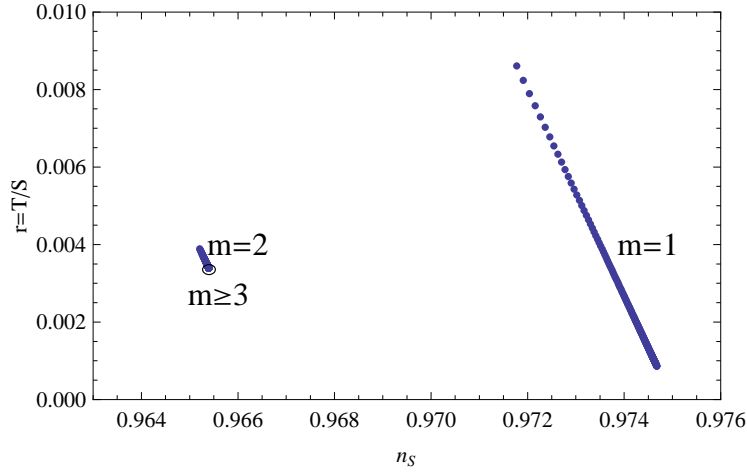}
\end{center}
\caption{The spectral index $n_s$ and the tensor-to-scalar perturbation ratio $r$
are depicted in one plot for various values of $a_0$ and the power of the
non-minimal coupling $m$ in $K(\phi)\sim \phi^m$. }
\label{rnplot}
\end{figure}

In fig.\ref{rnplot} we plotted the spectral index ($n_S$) and the tensor-to-scalar
perturbation ratio ($r$) for varying $a_0$ and fixed $N=60$. For $m=1$ and $m=2$,
the spectral index becomes larger but the tensor-to-scalar ratio becomes smaller.
For {\it large} $a_0\simeq 4\pi$,
the values of the spectral index and the tensor-to-scalar ratio are saturated to
$0.9745 (0.965)$ and $0.0007(0.003)$ for $m=1(m\geq 2)$, respectively. Notice that
when $m\geq 3$, the spectral index and $r$ are independent of $a_0$ and
given as $0.965$ and $0.003$, respectively. It is depicted by a circle at the tip
of the plot for $m=2$.

Another observable is the amplitude of the scalar perturbation.
\begin{eqnarray}
 \delta_{H}=\frac{\delta \rho}{\rho}\cong \frac{1}{5\sqrt{3}H}\frac{U^{3/2}}{\mpl U'}=1.91\times 10^{-5}.
\end{eqnarray}
This gives a constraint for the parameters
\begin{eqnarray}
 \frac{U}{\epsilon}=(0.027\mpl)^4.
\end{eqnarray}
In our model, the constraint is written, with dimensionless parameter
 $\lambda_0=\lambda \Mh^{2m-4}$, as follows.
\begin{eqnarray}
 \begin{array}{ll}
  \sqrt{\frac{\lambda_0}{a_0}}&\simeq 2.3\times 10^{-5},~~ (m=1)\\
  \sqrt{\frac{\lambda_0}{a_0^2(1+1/(6a_0))}}&\simeq 2.1\times 10^{-5},~~ (m=2)\\
  \sqrt{\frac{\lambda_0}{a_0^2}}&\simeq 1.5\times 10^{-5}\sqrt{m},~~ (m\ge 3).\\
 \end{array}
\label{condition2}
\end{eqnarray}
One should note that $\sqrt{\frac{\lambda_0}{a_0^2}}\sim 10^{-5}$ is
universally required to fit the observational data for general values of $m$.
However this is weird since the quartic coupling has to be extremely small
$\lambda \sim 10^{-10} a_0^2$ as we already noticed in the case with $m=2$.

\section{Conclusion}
In this paper, we examined the inflationary scenarios based on non-minimal coupling of a
scalar field with the Ricci scalar ($\sim K(\phi)R$). Taking conformal transformation, the resultant
scalar potential in the Einstein frame is shown to be flat at the large field limit if the condition
in eq.\ref{condition} is satisfied. This is one of the main result of this paper.

This class of models gets constraints from the recent cosmological observations
of the spectral index, tensor-to-scalar perturbation ratio as well as the amplitude of the potential.
We explicitly considered the monomial cases $K \sim \phi^m$ and found that this class of models are indeed good
agreement with the recent observational data: $n_S \simeq 0.964-0.975$ and $r \simeq 0.0007- 0.008$ for any value of $m$.
In fig.\ref{rnplot}, the predicted values for $n_S$ and $r$ are depicted. We explicitly read out the condition for
fitting the observed anisotropy of the CMBR by which essentially the amplitude of the potential is determined. The condition
does not look natural ($\sqrt{\lambda/a^2}\sim 10^{-5}$) at the first sight but we may understand this seemingly unnatural value
once we embed the theory in higher dimensional space-time. Details of higher dimensional embedding of the theory and
possible solution to the smallness of $\sqrt{\lambda/a^2}$ will be given in separate publication \cite{large volume}.

\acknowledgments
We are indebted to discussions with Jihn E. Kim, Alex Nielsen, Soo-Jong Rey, Zheng Sun and Takao Suyama.
SC is grateful to D. K. Hong, T. Shiromizu and S. Fujii for conversations and great hospitality during the stay at PNU and TITECH.
A part of this work was done during ``LHC phenomenology-focus week program" at IPMU (Dec. 17-21, 2007).
Another part of this work was done during ``Focus Program on Liouville, Integrability and Branes (4)'' at APCTP (Dec. 11-24, 2007).
The work of SY was supported in part by The Korea Research Foundation
and The Korean Federation of Science and Technology Societies Grant
funded by Korea Government (MOEHRD, Basic Research Promotion Fund),
KOFST BP Korea Program, KRF-2005-084-C00003, EU FP6 Marie Curie Research and Training Networks MRTN-CT-2004-512194 and HPRN-CT-2006-035863 through
MOST/KICOS.

\end{document}